\documentclass[sigconf,nonacm]{acmart}

\AtBeginDocument{%
  }
\setcopyright{none}
\usepackage{setspace}
\usepackage{graphicx}
\usepackage{multirow}
\graphicspath{{figures/}}

\AtEndPreamble{%
  \theoremstyle{acmdefinition}
  \newtheorem{assumption}[theorem]{Assumption}
  
  \theoremstyle{acmplain}
}

\begin{document}

\title{SPADE: Spatiotemporal Preference Alignment and Diversity Enhancement for Local Search}

\author{chenxu wang}
\email{975679801@qq.com}
\affiliation{%
  \institution{Meituan}
  \city{}
  \country{}
}

\author{Wenxuan Yang}
\email{wenxuanyang2469@gmail.com}
\affiliation{%
  \institution{Meituan}
  \city{}
  \country{}
}

\author{Yueqi Mou}
\email{mouyueqi@meituan.com}
\affiliation{%
  \institution{Meituan}
  \city{}
  \country{}
}

\author{Shuxian Bi}
\email{shuxianbi2000@gmail.com}
\affiliation{%
  \institution{Independent}
  \city{}
  \country{}
}

\author{Biao Tang}
\email{biao.tang@meituan.com}
\affiliation{%
  \institution{Meituan}
  \city{}
  \country{}
}

\author{Peng Yan}
\email{yanpeng04@meituan.com}
\affiliation{%
  \institution{Meituan}
  \city{}
  \country{}
}

\renewcommand{\shortauthors}{Wang et al.}

\begin{abstract}
    Proactive query recommendation aims to suggest relevant search queries to
    users on the homepage of local service apps.
    Unlike conventional recommendation over a fixed item set, this requires open-set generation from a vocabulary of millions, where outputs must simultaneously align with user preference, match fine-grained spatiotemporal context, cover long-tail intents, and provide diverse alternatives within a few candidate slots. However, existing approaches address these axes in isolation: standard tokenization suppresses long-tail queries, coarse location encoding misses fine-grained geographic demand, generic preference optimization fails to align with production impression-click distributions, and standard beam search collapses to near-duplicate outputs. To address these limitations, we present \textbf{SPADE} (\textbf{S}patiotemporal \textbf{P}reference
    \textbf{A}lignment and \textbf{D}iversity \textbf{E}nhancement), an
    end-to-end generative framework that jointly optimizes the generation space, contextual representation, preference distribution, and output slate quality through four targeted components:
    \emph{Objective-Aligned BPE (OA-BPE)} aligns subword merging with business
    value signals to boost long-tail query generation probability;
    \emph{Spatio-Temporal Adaptive Representation (STAR)} replaces coarse
    location tokens with continuous multi-scale encodings and soft gating for
    precise spatiotemporal grounding;
    \emph{MARS-DPO} (Multi-dimensional Adaptive Reward Sampling DPO) combines
    cross-domain pre-training with preference alignment on production
    impression-click data via composite rewards and adaptive hard negative
    sampling;
    and \emph{Sigmoid Diversity Beam Search (SDBS)} improves intent coverage
    without sacrificing recall.
    Offline experiments on large-scale production data show that SPADE outperforms the strongest generative
    baseline by +5.6\% HR@5, $-$40.3\% GeoC@5, and +1.9\% Exec@5.
    A one-week online A/B test confirms production impact: +1.88\% Module CTR,
    +0.177\% Search Volume, and +0.105\% Search DAU.
    \end{abstract}


\maketitle

 \begin{spacing}{0.98}
\section{Introduction}

\begin{figure}[!t]
  \centering
  \includegraphics[width=\linewidth]{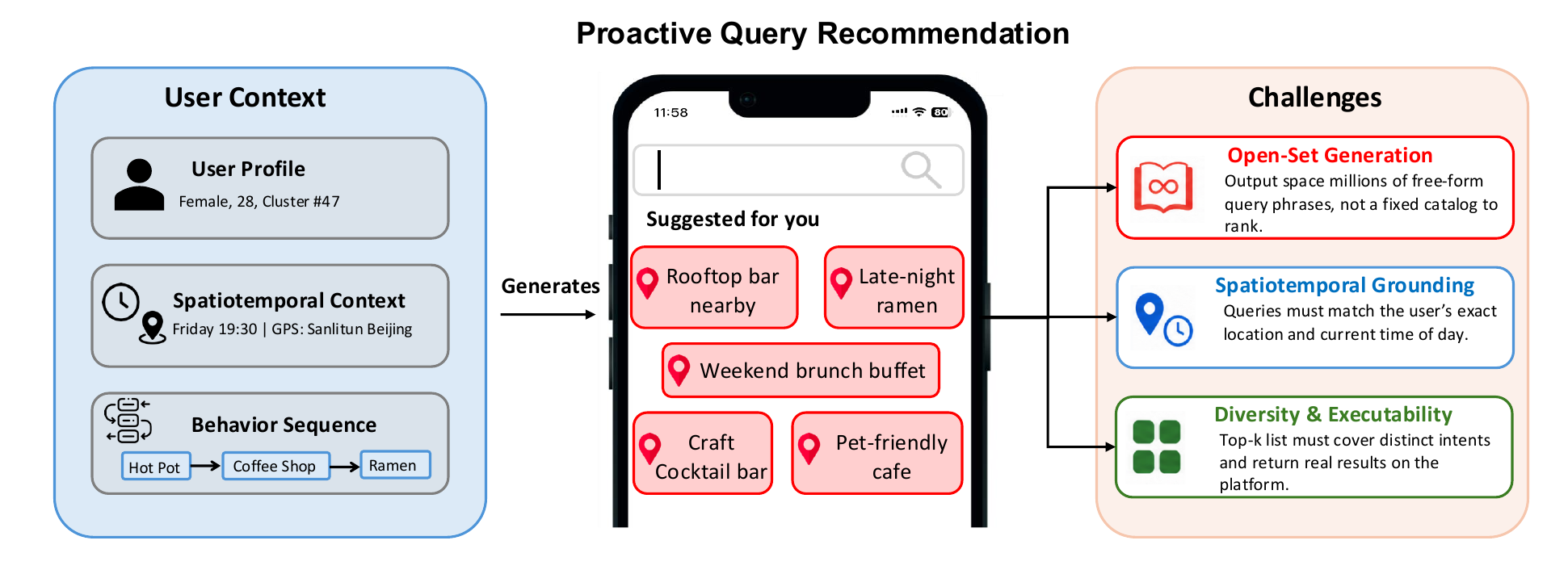}
  \Description{Motivational figure showing user context inputs on the left,
  a smartphone interface with suggested search queries in the middle, and
  three challenge callout boxes on the right.}
  \caption{Proactive query recommendation on a local service platform. Given
  the user's profile, spatiotemporal context, and behavior history, the system
  surfaces a ranked list of candidate search queries before the user types
  anything. The task is open-set generation: outputs are free-form phrases
  that must be spatiotemporally grounded, diverse, and executable by the
  platform's retrieval stack.}
  \label{fig:motivation}
\end{figure}

On local service platforms, proactive query recommendation presents a set of
candidate search intents the moment a user opens the
app~\cite{cite_onesug, cite_gqs}, as illustrated in
Figure~\ref{fig:motivation}. This interface accounts for a substantial share
of search clicks and downstream conversions, serving as the starting point
from app launch to completed transaction.

This task is fundamentally different from conventional recommendation.
Traditional recommenders assume a fixed, enumerable candidate
pool~\cite{cite_cf_rec, cite_sequential_rec}; proactive query recommendation
has no such inventory. The system must generate free-form phrases from a
vocabulary of millions, with no retrieval-based fallback. This makes it an open-set generation problem with a much larger and less constrained
output space. In local search the constraints tighten further. The available inputs are
limited to a user profile, current GPS location and time, and a short behavior
history. Yet each generated candidate must simultaneously satisfy four
requirements: (1)~long-tail intent coverage, surfacing commercially
valuable queries beyond popular head terms; (2)~spatiotemporal
grounding, matching the user's geographic and temporal context at fine
granularity; (3)~preference alignment, reflecting the user's genuine
interests even with sparse history; and (4)~decoding diversity,
offering distinct options within a small candidate slate where every item must
be executable by the downstream search engine. Any candidate that fails on
even one dimension wastes a display slot.

Existing methods each address a subset of these
requirements~\cite{cite_sr_survey, cite_genrec_survey}. Retrieval-based
approaches such as query log mining and query-flow
graphs~\cite{cite_qac, boldi2008queryflow} are bounded by the historical
candidate pool: they cannot cover long-tail intents absent from the logs and
struggle with cold-start users. Generative methods close part of this gap but
leave blind spots. AIGQ, OneSug, and CTR-GQS~\cite{cite_aigq, cite_onesug,
cite_gqs} concatenate location and time as text tokens in the prompt,
capturing only coarse city or time-slot-level patterns that cannot
distinguish a neighborhood center from its periphery. Generative POI methods
(OneLoc, Spacetime-GR, GeoGR~\cite{cite_oneloc, cite_spacetime_gr,
cite_geogr}) introduce geographic indexing but output structured POI
identifiers, not open-vocabulary queries. Across both families, standard BPE
and standard beam search are applied without adaptation: BPE optimizes for
corpus compression, putting high-value long-tail queries at a structural
disadvantage during autoregressive generation; beam search collapses the
top-$K$ list into near-duplicate head-intent
paraphrases~\cite{cite_beam_search_degeneration}. Few existing systems
consider more than one of these requirements jointly.

{\sloppy
To address these limitations, we propose \textbf{SPADE} (\textbf{S}patiotemporal \textbf{P}reference
\textbf{A}lignment and \textbf{D}iversity \textbf{E}nhancement), an
end-to-end generative framework that jointly optimizes four stages of the
generation pipeline---the generation space, contextual representation,
preference distribution, and output slate quality:
(1)~At the vocabulary level, \textbf{Objective-Aligned BPE (OA-BPE)}
incorporates business metrics into BPE merging~\cite{cite_bpe_orig} so
that high-value long-tail queries receive shorter tokenizations, injecting a
business-aligned generation prior before training begins.
(2)~At the input level, \textbf{Spatio-Temporal Adaptive Representation
(STAR)} replaces coarse discrete location tokens with multi-scale hexagonal
spatial and temporal encodings, injected through zero-initialized soft gates
that let language modeling stabilize before spatiotemporal signals are
gradually introduced.
(3)~At the training level, \textbf{MARS-DPO} (Multi-dimensional Adaptive
Reward Sampling DPO) first pre-trains on cross-domain search behavior, then
runs preference alignment on production impression-click data with composite
rewards (ranking position, retrieval quality, geographic proximity) and
adaptive hard negative sampling~\cite{cite_dpo_alignment}.
(4)~At the decoding level, \textbf{Sigmoid Diversity Beam Search (SDBS)}
applies a position-aware bonus to lower-ranked beams while leaving the top
beam undisturbed, improving diversity without sacrificing recall.
\par}

SPADE is, to our knowledge, one of the first generative query recommendation frameworks for local search. We summarize our contributions as follows:

\begin{itemize}
    \item \textbf{A unified, co-designed framework.} We propose SPADE, a
    generative query recommendation framework for local search in which the vocabulary, context encoding, preference alignment, and decoding stages are designed jointly: each stage operates on the output distribution of the previous one, so improvements compound rather than conflict.
    \item \textbf{Vocabulary and context encoding.} (i) \emph{OA-BPE},
    a business-aware tokenizer that shortens valuable long-tail queries,
    justified by a merge-preference analysis (Proposition~\ref{prop:merge});
    and (ii) \emph{STAR}, a multi-scale spatiotemporal encoding with
    soft-gated injection.
    \item \textbf{Preference alignment and decoding.} (i) \emph{MARS-DPO},
    a production-oriented preference alignment on real impression--click
    feedback; and (ii) \emph{SDBS}, a diversity-aware decoding strategy
    that improves slate quality while preserving relevance.
    \item \textbf{Experiments.} On large-scale production data, SPADE
    outperforms the strongest generative baseline (AIGQ) by +5.6\% HR@5,
    $-$40.3\% GeoC@5, and +1.9\% Exec@5, with gains reaching +13.5\% HR@5 on long-tail intents and $-$48.3\% GeoC@5 in geo-sparse regions. A one-week online A/B test confirms production impact: +1.88\% Module CTR, +0.177\% Search Volume, and +0.105\% Search DAU, with negligible latency overhead.
\end{itemize}

\section{Related Work}
\label{sec:related}

\subsection{Query Recommendation}
\label{subsec:query_rec}

Traditional query recommendation leverages co-occurrence statistics, session
transitions, and query-flow graphs to mine candidates from historical logs
\cite{baeza2004query, boldi2008queryflow}, but is limited to previously observed
patterns and struggles with cold-start users and long-tail intents. More
recently, the task has been reformulated as open-vocabulary text generation
\cite{geng2022p5, guo2025gqr}, with methods spanning local search
\cite{chen2026localsug}, conversational search \cite{cite_gqs}, e-commerce
\cite{cite_onesug}, and industrial pre-search recommendation \cite{cite_aigq}.
GQR \cite{guo2025gqr} aligns LLM generation with user preference via a
CTR-based listwise reward; AIGQ \cite{cite_aigq} combines listwise SFT with a
policy gradient objective and hybrid serving. Despite these advances, both target
generic or e-commerce settings and do not jointly model fine-grained geographic
context, temporal context, or business-aware tokenization for local search.
Data-centric approaches such as synthetic-prefix augmentation
\cite{rajan2025synthetic} address presentation bias in autocomplete but do not
close this gap.

Location-aware recommendation \cite{cheng2013mobile} and
spatial encoders \cite{mai2020space2vec} have long captured geographic relevance,
while spatiotemporal models such as STAN \cite{lin2021stsan} demonstrate the value
of temporal attention over mobility patterns. More recent generative local
methods---GNPR-SID, OneLoc, Spacetime-GR, GeoGR \cite{cite_gnpr_sid,
cite_oneloc, cite_spacetime_gr, cite_geogr}, and LGSID \cite{cite_lgsid}---verify
the importance of geographic and temporal grounding, but all optimize over POI
identifiers or semantic IDs rather than free-form executable queries. Generative
retrieval and sequential models such as TIGER \cite{cite_gen_rec}, P5
\cite{geng2022p5}, UniSRec \cite{unisrec2022}, and SASRec
\cite{cite_sequential_rec} similarly operate over fixed target spaces.

On vocabulary and alignment, standard BPE \cite{cite_bpe_orig} and SentencePiece
\cite{kudo2018sentencepiece} are optimized for compression rather than
recommendation value, and can systematically suppress rare but high-value phrases
\cite{cite_bpe_limitations}. Preference-learning methods including RLHF
\cite{ouyang2022rlhf}, DPO \cite{cite_dpo_alignment}, robust variants
\cite{ye2024rdpo}, and CausalDPO \cite{cite_causaldpo} have proven effective for
aligning generative models; listwise objectives \cite{ma2023listwise,
ma2024rankllama} further emphasize optimizing over candidate sets rather than
individual outputs. Effective local search query recommendation therefore requires
a unified treatment of open-vocabulary generation, spatiotemporal grounding,
business-aware tokenization, and preference alignment---a gap SPADE is designed
to close.

\subsection{Diversity in Generative Decoding}
\label{subsec:diversity}

Standard beam search collapses to near-duplicate hypotheses
\cite{holtzman2020curious}. Diverse Beam Search \cite{cite_beam_search_degeneration}
mitigates this with inter-group penalties; nucleus and adaptive sampling
\cite{holtzman2020curious, basu2021mirostat} broaden exploration; contrastive
decoding \cite{su2022contrastive, arias2024adaptive} penalizes degenerate
continuations; and MBR decoding \cite{jinnai2024mbr, eikema2020map} optimizes
set-level utility. D5P4 \cite{cite_d5p4} extends diversity control to discrete
diffusion via DPP-based beam selection, and ReCast \cite{cite_recast} addresses
sparse-hit RL by restoring learnability for all-zero rollout groups. These methods
are optimized for generic text quality or offline utility, however, and do not
capture the constraints of industrial query recommendation, where a short
candidate list must be jointly relevant, geographically grounded, and diverse
under tight latency budgets. SPADE addresses this with Sigmoid Diversity Beam
Search, a position-aware decoding strategy that improves intent coverage without
sacrificing recall or production viability.

\section{Method}
\label{sec:method}

\subsection{Task Definition}
\label{sec:task}

Given a user $u$ at time $t$, the system observes a user profile
$\mathbf{p}$, a behavioral sequence
$\mathcal{H} = (h_1, \ldots, h_n)$ of recent queries and visited POIs, and a
spatiotemporal context $(\mathbf{c}, t)$ with GPS coordinate
$\mathbf{c} = (\text{lon}, \text{lat})$. The goal is to generate $K$ candidate
queries $\mathcal{Q} = \{q_1, \ldots, q_K\}$ covering the four requirements
motivated above---long-tail coverage, spatiotemporal grounding, preference
alignment, and slate diversity. Because the output space is the platform's
full query vocabulary (millions of phrases), this is an open-set generation
problem.
SPADE addresses this with four components, each targeting a different stage of
the generation pipeline:
(1)~\textbf{OA-BPE} constructs a business-aligned vocabulary;
(2)~\textbf{STAR} encodes fine-grained spatiotemporal context as token
embeddings; (3)~a \textbf{two-stage training} pipeline pre-trains on
cross-domain search behavior and then runs preference alignment via MARS-DPO;
(4)~\textbf{Sigmoid Diversity Beam Search} produces a diverse candidate list
at inference time.

\subsection{Tokenizer: Objective-Aligned BPE (OA-BPE)}
\label{sec:oabpe}

To improve long-tail intent coverage, OA-BPE intervenes at the
vocabulary level. In a generative model, the probability of a token sequence
$y = (y_1, \ldots, y_T)$ is a product of per-token conditionals, each bounded
above by some $p_{\max} \le 1$, so longer tokenizations yield lower sequence
probabilities. Standard
BPE~\cite{cite_bpe_orig} merges token pairs by corpus frequency, optimizing
compression rather than business value. High-frequency but low-value tokens are
merged early, while rare but commercially valuable long-tail queries are
over-segmented and disadvantaged at generation time. Prior BPE modifications target
linguistic fidelity (e.g., preserving named entities or morphological
boundaries~\cite{cite_bpe_limitations}); OA-BPE instead incorporates business value
signals directly into the merge criterion, giving high-value queries shorter
tokenizations as a generation prior with no additional training cost.

OA-BPE replaces the standard merge score with a business-value-weighted count:
\begin{equation}
  \text{score}_{\text{OA}}(a, b)
    = \sum_{s \in \mathcal{C}} w(s) \cdot \text{count}_s(a, b),
  \label{eq:oabpe-score}
\end{equation}
where $\mathcal{C}$ is the training corpus (the set of all platform queries),
$s$ denotes an individual query in $\mathcal{C}$, $\text{count}_s(a,b)$
is the occurrence count of token pair $(a,b)$ in $s$, and $w(s)$ is a
business value weight that aggregates platform-specific engagement and
conversion signals. Standard BPE is the special case $w(s) \equiv 1$;
complexity is identical ($O(N \cdot V)$).
In our experiments we instantiate $w(s)$ as a weighted combination of
click-through rate (CTR), conversion rate (CVR), and a module-exposure normalization term
$1 - 1/(\log(\text{PV}(s)+1)+1)$, which compresses head-query dominance while
preserving meaningful volume differences in the long tail.

Proposition~\ref{prop:merge} quantifies how much frequency disadvantage the
value weighting can overcome.

\begin{proposition}[Value-weighted merge preference]
\label{prop:merge}
Let pairs $(a,b)$ and $(a',b')$ have corpus counts and count-weighted mean
business values
\[
c = {\textstyle\sum_s}\,\text{count}_s(a,b), \qquad
\bar{w} = c^{-1}{\textstyle\sum_s}\, w(s)\,\text{count}_s(a,b),
\]
with $c'$ and $\bar{w}'$ defined analogously. Standard BPE merges $(a,b)$
first iff $c > c'$, whereas OA-BPE does so iff $\bar{w} c > \bar{w}' c'$.
Hence OA-BPE merges the higher-value pair first even when it is up to
$\bar{w}/\bar{w}'$ times rarer.
\end{proposition}

\begin{proof}
By definition of $\bar{w}$, Eq.~\eqref{eq:oabpe-score} gives
$\text{score}_{\text{OA}}(a,b) = \bar{w} c$, and standard BPE is the case
$w \equiv 1$, i.e., $\bar{w}=\bar{w}'=1$. Comparing the two scores,
$\bar{w}c > \bar{w}'c'$ holds whenever $c/c' > \bar{w}'/\bar{w}$.
\end{proof}

\noindent
Section~\ref{subsec:component_analysis} measures both the tokenization and
generation-probability effects: tail-query median token length drops from $7.2$
to $5.9$ while head queries are left essentially unchanged.

\subsection{Input Encoding: Spatio-Temporal Adaptive Representation (STAR)}
\label{sec:star}

With the OA-BPE vocabulary, the input sequence is organized into three parts:
\begin{equation}
  \mathbf{x} = \bigl[
    \underbrace{p_1, \ldots, p_m}_{\text{user}},\;
    \underbrace{c_1, \ldots, c_l}_{\text{context}},\;
    \underbrace{h_1, \ldots, h_n}_{\text{history}}
  \bigr],
  \label{eq:input}
\end{equation}
where user tokens encode profile attributes (demographics, cluster
membership, etc.), context tokens capture coarse-grained situational
signals (city, district, day of week, time slot, entry scenario, device type,
etc.), and history is the concatenated text of recent search queries
and visited POI names. The first $K$ positions within the context segment are
reserved as placeholders for STAR, which injects fine-grained continuous
spatiotemporal embeddings to complement the discrete context tokens.

SPADE uses an encoder--decoder Transformer. Token embeddings are scaled by
$\sqrt{d_{\text{model}}}$, and STAR injects spatiotemporal embeddings additively
into the first $K$ positions. Each encoder layer uses a
Sandwich Norm structure~\cite{ding2021cogview} combining gated self-attention with
per-head QKNorm~\cite{henry2020qknorm} and a SwiGLU FFN~\cite{shazeer2020glu};
the decoder adds an encoder--decoder cross-attention sublayer.

\begin{figure}[tb]
  \centering
  \includegraphics[width=\linewidth]{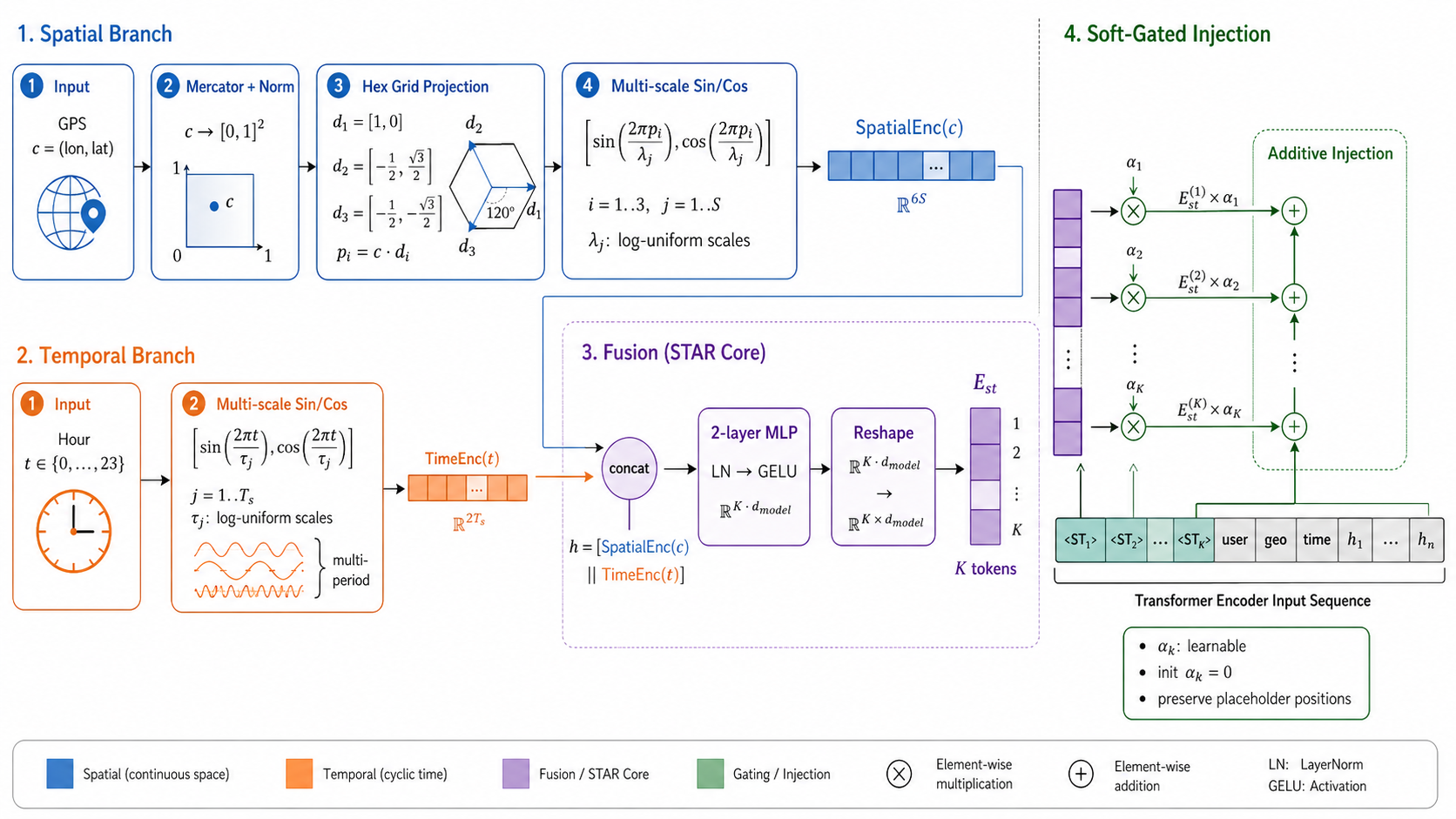}
  \Description{Architecture diagram of the STAR module showing spatial and
  temporal encoding branches merging into an MLP, with soft-gated additive
  injection into encoder placeholder positions.}
  \caption{STAR module architecture. Spatial and temporal encodings are
  produced by multi-scale sinusoidal branches, fused via a two-layer MLP,
  and injected additively into the encoder's placeholder positions through
  independent learnable soft gates $\alpha_k$ (initialized to zero).}
  \label{fig:star_module}
\end{figure}

Discrete context tokens (e.g., city IDs) have a fixed precision ceiling and
cannot express distance relationships between continuous coordinates. STAR encodes
exact GPS coordinates and the hour of day as $K$ token-level embeddings injected
additively into the encoder input, complementing the discrete channel.

\subsubsection{Spatiotemporal Encoding}
Following Space2Vec~\cite{mai2020space2vec}, we encode coordinates with
multi-scale sinusoidal functions along three hexagonal axes spaced $120^\circ$
apart, providing isotropic spatial resolution. For normalized coordinates
$\mathbf{c} \in [0,1]^2$ (Web Mercator projection), the scalar projection onto
direction $i$ is $p_i = \mathbf{c} \cdot \mathbf{d}_i$, and the spatial encoding
is:
\begin{equation}
  \text{SpatialEnc}(\mathbf{c})
    = \bigoplus_{i=1}^{3} \bigoplus_{j=1}^{S}
      \Bigl[\sin\!\Bigl(\tfrac{2\pi p_i}{\lambda_j}\Bigr),\,
            \cos\!\Bigl(\tfrac{2\pi p_i}{\lambda_j}\Bigr)\Bigr],
  \label{eq:spatial-enc}
\end{equation}
where the $S$ spatial scales $\lambda_1, \ldots, \lambda_S$ are log-uniformly
distributed over $[\lambda_{\min}, \lambda_{\max}]$, covering fine-grained
commercial districts to city-level regions.

Temporal encoding follows the same multi-scale sinusoidal scheme for the hour of
day $t$:
\begin{equation}
  \text{TimeEnc}(t)
    = \bigoplus_{j=1}^{S_t}
      \Bigl[\sin\!\Bigl(\tfrac{2\pi t}{\tau_j}\Bigr),\,
            \cos\!\Bigl(\tfrac{2\pi t}{\tau_j}\Bigr)\Bigr],
  \label{eq:time-enc}
\end{equation}
with $S_t$ temporal scales $\tau_1, \ldots, \tau_{S_t}$ log-uniformly
distributed over $[\tau_{\min}, \tau_{\max}]$, capturing intra-hour to diurnal
patterns.

\subsubsection{Fusion and Soft-Gated Injection}
The spatial and temporal encodings are concatenated and projected through a
two-layer MLP, then reshaped into $K$ token embeddings:
\begin{align}
  \mathbf{h} &= \bigl[\text{SpatialEnc}(\mathbf{c}) \;\|\; \text{TimeEnc}(t)\bigr], \label{eq:star-concat} \\
  \mathbf{E}_{\text{st}} &= \text{Reshape}\bigl(
      W_2 \cdot \text{GELU}(W_1 \mathbf{h} + b_1) + b_2
    \bigr)
    \in \mathbb{R}^{K \times d_{\text{model}}}.
  \label{eq:star-mlp}
\end{align}
Using $K > 1$ allows different tokens to specialize in different spatiotemporal
granularities.

Each token has an independent learnable gate $\alpha_k$, initialized to zero:
\begin{equation}
  \mathbf{E}[:, k, :] \leftarrow \mathbf{E}[:, k, :] + \alpha_k \cdot E_{\text{st}}^{(k)},
  \quad k = 1, \ldots, K.
  \label{eq:soft-gate}
\end{equation}
With $\alpha_k = 0$ at initialization, STAR has no effect early in training,
letting language modeling capacity stabilize first; as training proceeds,
$\alpha_k$ grows adaptively. Compared with directly replacing the placeholder
embeddings, additive injection retains the original positional encoding and
thus preserves more input information (this design choice is validated by
ablation in Section~\ref{subsec:ablation}).

\subsection{Training: Alignment via MARS-DPO}
\label{sec:training}

Training proceeds in two stages: cross-domain pre-training builds broad
intent understanding, and MARS-DPO then aligns generation with real user
feedback on production impression--click data.

\subsubsection{Stage 1: Cross-Domain Pre-training}
The first stage trains SPADE on a Next Search Query Prediction task over
cross-domain user search logs covering multiple verticals (restaurants,
entertainment, lifestyle services, etc.). The loss is temperature-scaled
cross-entropy:
\begin{equation}
  \mathcal{L}_{\text{pre}}
    = -\frac{1}{|T|} \sum_{t \in T} \log
      \frac{\exp(z_{y_t} / \tau)}{\sum_{v} \exp(z_v / \tau)},
  \label{eq:pretrain-loss}
\end{equation}
where $\tau > 1$ smooths the prediction distribution, encouraging diverse
generation paths. Input features are randomly masked with probability
$p_{\text{mask}}$ to improve robustness to sparse inputs at inference time.

\subsubsection{Stage 2: Preference Alignment via MARS-DPO}
The second stage aligns the model with real user behavior using \textbf{MARS-DPO}
(Multi-dimensional Adaptive Reward Sampling DPO), trained on impression--click
data from the query recommendation module. MARS-DPO adapts DPO to this
production setting along three axes: online candidate reuse,
multi-dimensional rewards, and adaptive hard negatives. Unlike conventional
DPO training, which optimizes over static, pre-collected preference pairs,
MARS-DPO constructs its pairs directly from each request's online recall
list---matching the serving traffic distribution at no additional rollout
cost---and dynamically selects hard negatives from the same list.

\vspace{2pt}
\noindent
\textbf{Multi-dimensional reward.} Each candidate $q$ receives a composite reward:
\begin{equation}
  R(q) = w_1 \cdot R_{\text{rank}}(q) + w_2 \cdot R_{\text{qual}}(q) + w_3 \cdot R_{\text{dist}}(q),
  \label{eq:reward}
\end{equation}
where $w_1, w_2, w_3$ are tuned on the validation set.

\textit{(a) Rank reward} uses the online ranker's position as a relevance signal:
$R_{\text{rank}}(q) = -\log(\text{rank}(q) + 1)$.

\textit{(b) Quality reward} assesses executability: given $q$'s top retrieved
documents $\{d_1, \ldots, d_n\}$, we query Qwen3.5-122B-A10B via batch inference
and use the logit probability ratio of the ``yes'' and ``no'' tokens as a
continuous reward:
\begin{equation}
  R_{\text{qual}}(q)
    = \frac{P_{\text{LLM}}(\texttt{"yes"} \mid \text{prompt}(q, \{d_i\}))}
           {P_{\text{LLM}}(\texttt{"yes"} \mid \cdot) + P_{\text{LLM}}(\texttt{"no"} \mid \cdot)}.
  \label{eq:qual-reward}
\end{equation}
This provides smoother gradients than hard binary
judgments~\cite{zhang2025qwen3reranker}. The reward is computed offline via batch
inference over the training corpus.

\textit{(c) Distance reward} penalizes geographically distant results:
\begin{equation}
  R_{\text{dist}}(q)
    = 1 - \text{MinMaxNorm}\!\bigl(\log(d_{\min}(q) + 1)\bigr),
  \label{eq:dist-reward}
\end{equation}
where $d_{\min}(q)$ is the distance from the user to the nearest POI in $q$'s
top retrieved results.

\vspace{2pt}
\noindent
\textbf{Adaptive hard negative selection.} The positive sample is
$q^+ = \arg\max_{q \in \mathcal{C}} R(q)$. We rank all remaining candidates by
reward gap $\Delta_i = R(q^+) - R(q_i)$ and select the top-$K$ with smallest
$\Delta_i$ as the hard negative set $\mathcal{N}_K$. Each negative is then
weighted by its softmax-normalized proximity:
\begin{equation}
  w(q_i^-) = \frac{\exp(-\eta \cdot \Delta_i)}{\sum_{q_j \in \mathcal{N}_K} \exp(-\eta \cdot \Delta_j)},
  \label{eq:hard-neg}
\end{equation}
where $\eta$ controls the concentration of weights on the hardest negatives.
For each positive--negative pair $(q^+, q_i^-)$, the DPO objective is:
\begin{equation}
\begin{aligned}
  \mathcal{L}_{\text{DPO}}(q^+, q_i^-)
    = -\log \sigma\!\Biggl(\beta \cdot \biggl[&
        \log \frac{\pi_\theta(q^+ \mid x)}{\pi_{\text{ref}}(q^+ \mid x)}\\
        -{}& \log \frac{\pi_\theta(q_i^- \mid x)}{\pi_{\text{ref}}(q_i^- \mid x)}
      \biggr]\Biggr).
\end{aligned}
  \label{eq:dpo}
\end{equation}
The multi-negative loss aggregates over $\mathcal{N}_K$:
\begin{equation}
\begin{aligned}
  \mathcal{L}_{\text{multi}}
    &= \sum_{q_i \in \mathcal{N}_K} w(q_i^-)\\
    &\quad {}\cdot \mathcal{L}_{\text{DPO}}(q^+, q_i^-).
\end{aligned}
  \label{eq:multi-neg}
\end{equation}

\vspace{2pt}
\noindent
\textbf{SFT anchoring.} Pure DPO can cause both positive and negative
probabilities to decrease. We add an SFT auxiliary loss as an absolute anchor:
\begin{equation}
  \mathcal{L}_{\text{SFT}} = -\log P_\theta(q^+ \mid x), \quad
  \mathcal{L}_{\text{sample}} = \mathcal{L}_{\text{multi}} + \lambda_{\text{sft}} \cdot \mathcal{L}_{\text{SFT}}.
  \label{eq:sft-anchor}
\end{equation}

\vspace{2pt}
\noindent
\textbf{Impression feedback weighting.} Training samples carry varying reliability.
We assign sample-level weights $\omega_{\text{imp}} < \omega_{\text{clk}} <
\omega_{\text{cvr}}$ to impressions without clicks, with clicks, and with
downstream conversions:
\begin{equation}
  \mathcal{L}_{\text{total}}
    = \frac{1}{|\mathcal{D}|}
      \sum_{(x, q^+, q^-) \in \mathcal{D}} \omega(x) \cdot \mathcal{L}_{\text{sample}}(x, q^+, q^-).
  \label{eq:total-loss}
\end{equation}

MARS-DPO adopts pointwise alignment: impression--click logs provide
per-candidate signals, and slate-level diversity is delegated to the decoding
stage (Section~\ref{sec:sdbs}).

\subsection{Decoding: Sigmoid Diversity Beam Search (SDBS)}
\label{sec:sdbs}

To improve decoding diversity, we replace standard beam search with a
position-aware diversity bonus at inference time.

\begin{figure}[tb]
  \centering
  \includegraphics[width=\linewidth]{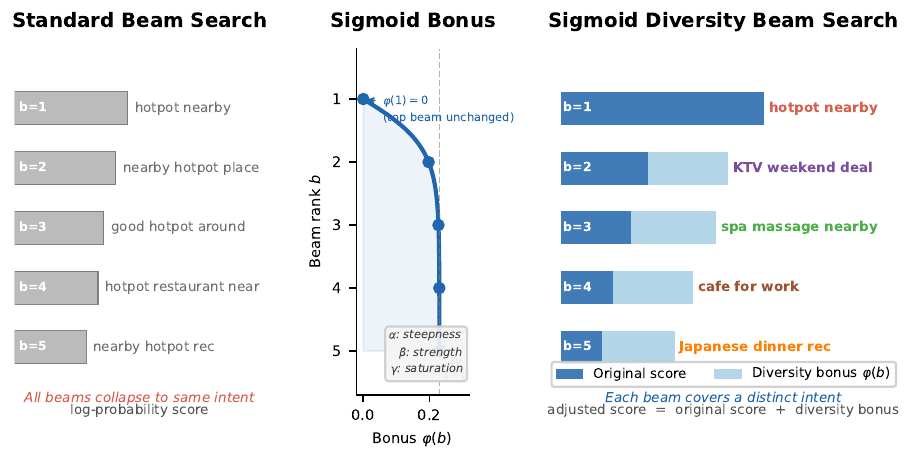}
  \Description{Three-panel illustration of Sigmoid Diversity Beam Search.
  Left: standard beam search where all five beams collapse to paraphrases
  of the same intent. Center: the sigmoid bonus curve showing how the bonus
  increases with beam rank while remaining zero for beam 1. Right: SDBS
  output where each beam covers a distinct intent with stacked original
  score and diversity bonus segments.}
  \caption{Sigmoid Diversity Beam Search. Standard beam search collapses
  all beams to paraphrases of one intent (left). The sigmoid bonus
  $\varphi(b)$ (center) is zero for the top beam and increases with rank,
  preserving the best candidate while boosting lower-ranked beams. With
  SDBS (right), each beam covers a distinct intent; bar segments show the
  original score (dark) and diversity bonus (light).}
  \label{fig:sdbs}
\end{figure}

Standard beam search ranks candidates by sequence probability alone, causing all
$B$ outputs to collapse to near-duplicate queries (e.g., ``hot pot restaurant'',
``hot pot delivery'', ``hot pot nearby''), as illustrated in
Figure~\ref{fig:sdbs} (left). We address this with a position-aware
diversity bonus $\phi(b)$ added to beam~$b$'s score at each decoding step
(Figure~\ref{fig:sdbs}, center and right).

Diverse Beam Search (DBS)~\cite{cite_beam_search_degeneration} addresses
mode collapse by partitioning beams into groups and applying Hamming-distance
penalties between groups. This inter-group penalty operates on surface-level
token overlap, which can penalize legitimate variations (e.g., ``hot pot
restaurant'' vs.\ ``hot pot nearby'') while missing semantically redundant
pairs that share few tokens. DBS also perturbs all beams, including
the top beam, risking degradation of the highest-quality output.

Our approach differs in two ways. First, SDBS uses a position-based
bonus rather than a token-overlap penalty: it encourages lower-ranked beams
to survive the pruning step without requiring explicit pairwise comparisons,
making it both simpler ($O(B)$ vs.\ $O(B^2)$) and agnostic to surface form.
Second, the bonus is designed to satisfy four properties:
(i)~zero at $b=0$ so beam~0 is never perturbed and the highest-quality
output is preserved; (ii)~strictly increasing in $b$ so lower-ranked beams
receive progressively larger incentives to diverge; (iii)~concave ($\phi''(b)\leq 0$) so the marginal bonus diminishes,
preventing excessive perturbation of low-ranked beams; and (iv)~bounded above so diversity
never overrides model confidence.
Properties~(ii)--(iv) are not arbitrary design choices. In the theoretical
analysis below we model beam scores as Zipf-distributed
(Assumption~\ref{ass:zipf}) and show that the budget-optimal promotion shape
for rescuing pruned lower-ranked candidates is \emph{necessarily} increasing
and concave (Theorem~\ref{thm:rescue}): the score deficit of a pruned
candidate grows only logarithmically in rank, so a concave bonus tracks it
with minimal budget, while diminishing marginal coverage gains favor rescuing
contiguous low ranks over distant ones. Boundedness (iv) is what ties the
promotion to top-1 confidence preservation
(Corollary~\ref{cor:threshold}).
A shifted sigmoid satisfies all four properties:
\begin{equation}
  \phi(b) = \beta\!\left(\tfrac{1}{1+\gamma e^{-\alpha b}} - \tfrac{1}{1+\gamma}\right),
  \label{eq:sigmoid-penalty}
\end{equation}
with $\alpha > 0$ (\emph{steepness}: how rapidly the bonus grows with beam
rank), $\beta > 0$ (\emph{strength}: overall magnitude of the diversity bonus),
and $\gamma \in (0, 1]$ (\emph{saturation}: the fraction of $\beta$ reachable
at the asymptote, $\lim_{b\to\infty}\phi(b)=\beta\gamma/(1+\gamma)$).
The constraint $\gamma \leq 1$ additionally ensures global concavity on
$b \geq 0$: the inflection point of the sigmoid lies at
$b^{*} = \ln(\gamma)/\alpha \leq 0$, so $\phi''(b) \leq 0$ for all $b \geq 0$.
Algorithm~1 gives the complete procedure.

\begin{figure}[!tb]
\begin{minipage}{\linewidth}
\small\raggedright
\hrule\vspace{3pt}
\textbf{Algorithm 1:} Sigmoid Diversity Beam Search
\vspace{2pt}\hrule\vspace{4pt}
\textbf{Input:} beam size $B$; sigmoid params $\alpha,\beta,\gamma$ (Eq.~\ref{eq:sigmoid-penalty})\\[2pt]
Init $B$ beams: $S_b\!\leftarrow\!0$,\ $\mathbf{y}_b\!\leftarrow\!\langle\texttt{bos}\rangle$\\[2pt]
\textbf{for} each step $t = 1, 2, \ldots$ \textbf{do}\\[1pt]
\hspace*{1.5em}\textbf{for} $b = 0,\ldots,B{-}1$: expand $\mathbf{y}_b$ to top-$2B$ next tokens\\[1pt]
\hspace*{1.5em}Merge; score each $(b,v)$ pair:\\[1pt]
\hspace*{3em}$\tilde{S}_{b,v} = S_b + \log P(v\!\mid\!\mathbf{y}_b) + \phi(b)$\\[1pt]
\hspace*{1.5em}Keep top-$B$ $(b,v)$ pairs as next beams\\[2pt]
\textbf{return} top-$B$ finished sequences
\vspace{3pt}\hrule
\end{minipage}
\end{figure}

\vspace{2pt}
\noindent
\textbf{Design analysis.}
To guide the choice of $\phi$, we analyze SDBS under a simplified
single-step model of beam scores: Theorem~\ref{thm:rescue} gives a
shape-independent limit on what bounded rank-based promotion can rescue,
together with the budget-optimal rescue shape---from which the concavity of
$\phi$ is \emph{derived} rather than assumed---and
Corollary~\ref{cor:threshold} identifies the budget scale at which the
top-1 margin is necessarily exceeded. The model abstracts away token-level
autoregressive dynamics; its role is to provide intuition for parameter
selection, while sequence-level behavior is validated empirically
(Section~\ref{subsec:ablation}). Write
$\phi_{\max} = \sup_{b\ge 0}\phi(b) = \beta\gamma/(1+\gamma)$.

\begin{assumption}[Zipfian beam scores]
\label{ass:zipf}
At a given decoding step, the candidate sequence probabilities decay with
rank $r$ as $P_r = C(r+1)^{-\zeta}$, $\zeta > 0$, where $r = 0$ is the
top-ranked candidate; equivalently, $S_r = \log C - \zeta\log(r+1)$.
\end{assumption}

Two consequences are used below: the top-1 confidence margin is a constant,
$g = S_0 - S_1 = \zeta\log 2$; and the deficit of a pruned candidate
$r \ge B$ relative to the boundary beam grows only logarithmically,
\begin{equation}
  D(r) \;=\; S_{B-1} - S_r \;=\; \zeta \log \tfrac{r+1}{B},
  \label{eq:deficit}
\end{equation}
Assumption~\ref{ass:zipf} deliberately abstracts away the token-level,
autoregressive nature of beam search, in which hypotheses are re-sorted at
every step over sequence probabilities rather than a static rank
distribution. The results below are structural statements about this
idealized regime, intended as design intuition for the shape and scale of
$\phi$, not as guarantees on generated sequences.

\noindent
A pruned candidate $r \ge B$ outranks the boundary beam iff
$S_r + \phi(r) \ge S_{B-1} + \phi(B-1)$, i.e., iff its bonus
\emph{differential} covers the deficit (note $\phi(B{-}1) > 0$):
\begin{equation}
  \phi(r) - \phi(B-1) \;\ge\; D(r).
  \label{eq:survival}
\end{equation}
This is a single-step, boundary-beam benchmark; intermediate ranks also
collect bonuses, so exact survival requires entering the top-$B$ bonused
scores. Since additional distinct intents yield diminishing marginal
coverage, the promotion should rescue a \emph{contiguous} window $[B, h]$
rather than scattered high ranks. We ask how far a budget $M$ can reach, and
which shape spends it most economically.

\begin{theorem}[Rescue horizon and optimal shape]
\label{thm:rescue}
Fix a decoding step, suppose Assumption~\ref{ass:zipf} holds, let $\phi$
satisfy (i)--(iv) with $\sup_{b \ge 0}\phi(b) = M$, and treat the rank domain
as continuous.
(a)~No promotion satisfying (i)--(iv)---of any shape---can rescue a rank
beyond
\begin{equation}
  \bar{r}(M) \;=\; B\,e^{(M - \phi(B-1))/\zeta} - 1 \;\le\; B\,e^{M/\zeta} - 1.
  \label{eq:horizon}
\end{equation}
(b)~If every rank of a contiguous window $[B, h]$
satisfies~\eqref{eq:survival}, then
\begin{align}
  M &\ge \zeta\Big[\log\tfrac{h+1}{B} + \tfrac{B-1}{B}\Big],
  \quad\text{equivalently} \notag\\
  h &\le h^{*}(M) = B\,e^{M/\zeta - (B-1)/B} - 1,
  \label{eq:window}
\end{align}
and the bound is tight: it is attained by the $C^1$
\emph{linear-to-logarithmic} shape
\begin{equation}
  \tilde{\phi}(r) =
  \begin{cases}
    \tfrac{\zeta}{B}\, r, & 0 \le r \le B-1, \\[4pt]
    \zeta \log \tfrac{r+1}{B} + \zeta\,\tfrac{B-1}{B}, & r \ge B-1,
  \end{cases}
  \label{eq:optimal-shape}
\end{equation}
truncated at $\phi^{*}(r) = \min(\tilde{\phi}(r), M)$, which satisfies
(i)--(iv) and meets \eqref{eq:survival} with equality throughout the
unsaturated region.
\end{theorem}

\begin{proof}
(a)~By~\eqref{eq:survival}, $\zeta\log\tfrac{r+1}{B} \le \phi(r) -
\phi(B{-}1) \le M - \phi(B{-}1) \le M$; solving for $r$ gives both bounds.
(b)~Write $c = \phi(B-1)$. Since $\phi(r) \ge D(r) + c$ on $[B{-}1, h]$ with
equality at $r = B{-}1$, the right derivatives compare as
$\phi'(B{-}1^{+}) \ge D'(B{-}1) = \zeta/B$; concavity then forces the secant
slope $c/(B{-}1) = (\phi(B{-}1) - \phi(0))/(B{-}1) \ge \zeta/B$, i.e.,
$c \ge \zeta(B{-}1)/B$, and $M \ge \phi(h) \ge D(h) + c$
yields~\eqref{eq:window}.
\end{proof}

\noindent
Part~(a) implies that SDBS enlarges the effective search width by at most
$(\bar r + 1)/B \le e^{M/\zeta}$ relative to standard beam search.

\noindent
The optimal shape ramps linearly up to the boundary beam and then tracks the
deficit logarithmically: increments are front-loaded, since a flat-then-steep
shape wastes budget on distant ranks. The shifted
sigmoid of Eq.~\eqref{eq:sigmoid-penalty} is such a relaxation of
$\phi^{*}$: it shares the front-loaded concave ramp, and its
saturation---the price of boundedness---caps the rescue window at
$h^{*}(\phi_{\max})$. Two design rules follow: \emph{critical steepness},
$\phi(B) - \phi(B-1) \ge \zeta\log\tfrac{B+1}{B} \approx \zeta/B$ (otherwise
no rank is rescued at all); and \emph{slope matching},
$\phi'(0) = \alpha\beta\gamma/(1+\gamma)^2 \approx \zeta/B$, a principled
initialization for $(\alpha, \beta, \gamma)$.

\begin{corollary}[Structural relevance--diversity threshold]
\label{cor:threshold}
Under Assumption~\ref{ass:zipf}, any promotion that rescues even
the first pruned rank ($h = B$) requires budget
$M \ge \zeta\bigl[\log\tfrac{B+1}{B} + \tfrac{B-1}{B}\bigr] > \zeta\log 2
= g$; its budget necessarily exceeds the top-1 confidence margin. Rescue
and exact top-1 preservation are therefore in tension in this model,
suggesting that the relevance--diversity trade-off has a structural origin
rather than being a tuning artifact.
\end{corollary}

\noindent
This follows by setting $h = B$ in~\eqref{eq:window}: the bracket equals
$\log(1{+}1/B) + (B{-}1)/B$, which exceeds $\log 2$ for every $B \ge 2$.
Corollary~\ref{cor:threshold} predicts a plateau-then-degrade relevance curve
as the budget grows---precisely the behavior in
Figure~\ref{fig:diversity}, where HR@5 stays flat up to the production
operating point ($\beta = 0.4$) and degrades only beyond it. With scores
fluctuating around the Zipf trend, rescues occur probabilistically below the
threshold, consistent with the diversity gains at unchanged HR@5 in
Section~\ref{subsec:ablation}.
\noindent
These single-step results do not translate directly to sequence-level
decoding: hypotheses are re-sorted at every step, and over $T$ steps the
bonus accumulates to at most $T\phi_{\max}$. One property does hold step by
step: since beam $0$ is unbonused while every rival gains at most
$\phi_{\max}$, the top-ranked continuation coincides with standard beam
search whenever the leading margin $s_0 - \max_{b \ge 1} s_b$ exceeds
$\phi_{\max}$. Whether the model operates in this regime in practice is an
empirical question; Section~\ref{subsec:ablation} confirms it, with SDBS
leaving HR@5 unchanged relative to standard beam search while reducing
IDiv@5 by 8.7\%. Since the bonus is rank-asymmetric (zero at $b = 0$), it
also partially counteracts the length penalty that OA-BPE reduces at the
vocabulary level (Section~\ref{sec:oabpe}).

\section{Experiments}

Our experiments answer four questions.
\textbf{RQ1}: Does SPADE outperform strong industrial and generative baselines?
\textbf{RQ2}: How much does each component contribute?
\textbf{RQ3}: Where are the gains largest: cold-start users, long-tail intents,
sparse regions, or peak hours?
\textbf{RQ4}: Do offline gains hold in production?

\subsection{Experimental Setup}
\label{subsec:setup}

\textbf{Dataset and split.} We randomly sample 100K users from the production logs of a large-scale local
service platform; both training stages operate on this shared user pool.
Stage~1 (cross-domain pre-training) collects the sampled users' search logs
across all verticals (restaurants, entertainment, lifestyle services, etc.)
over 30 consecutive days. Stage~2 (MARS-DPO alignment) uses impression--click
pairs from the query suggestion module for the same users over the same
30-day window. Each instance
contains the user profile, a recent behavior sequence (truncated to 30 interactions),
GPS-level geographic context, and temporal context (hour-of-day, day-of-week,
holiday flag). We use a strict chronological split: the model is trained on the
30-day window, validated on the immediately following day, and tested on the day
after that. The test-set labels are collected from the held-out day, and no evaluation-day
instances are used in preference-pair construction or candidate generation.
Table~\ref{tab:data} summarizes the corpus statistics.

\begin{table}[t]
  \centering
  \caption{Dataset statistics.}
  \label{tab:data}
  \small
  \begin{tabular}{lr}
    \toprule
    Statistic & Value \\
    \midrule
    \# Sampled users           & 100K   \\
    \# Unique queries          & 1.48M  \\
    \# Cities                  & 2,608  \\
    Max behavior seq.\ length  & 30     \\
    Avg.\ behavior seq.\ length & 27.7  \\
    \midrule
    Stage 1 samples (pre-training)  & 7.79M \\
    Stage 2 samples (alignment)     & 3.08M \\
    \bottomrule
  \end{tabular}
\end{table}

\vspace{2pt}
\noindent
\textbf{Offline metrics.} Following prior work~\cite{cite_gqs, cite_onesug, cite_aigq}, we evaluate four
aspects of quality. \emph{Relevance}: HR@$K$, MRR@$K$, and SemSim@$K$ measure
exact-match recall, rank quality, and embedding-space proximity to the label
query, respectively. \emph{Diversity}: Intra-List Diversity (IDiv@$K$,
mean pairwise cosine similarity within the top-$K$ list; lower is more diverse).
\emph{Local grounding}: Geo-Consistency (GeoC@$K$) is the mean minimum distance
(km) from the user's GPS location to the nearest POI in the retrieved results for
each generated query; lower values indicate better geographic alignment.
\emph{Executability}: Search Executability (Exec@$K$), an LLM-as-judge score
(0--1) measuring whether each generated query returns suitable results in the
platform's retrieval system. To avoid correlated bias with the training reward
$R_{\text{qual}}$, Exec@$K$ uses a separate judge model (Qwen3-32B) with an
independently designed prompt.
We report results at $K=5$, which matches the production display slot count.

\vspace{2pt}
\noindent
\textbf{Implementation details.}
The backbone is a 6-layer Transformer decoder with hidden size 512 and 8
attention heads. Stage~1 pre-training uses a learning rate of 5e-5, batch size
256, temperature $\tau=2.0$, and random feature masking probability
$p_{\text{mask}}=0.15$, trained for 3 epochs. Stage~2 MARS-DPO uses a learning
rate of 1e-6, batch size 256, DPO temperature $\beta=0.1$, and SFT anchor
weight $\lambda_{\text{sft}}=0.5$, trained for 1 epoch. The composite reward
weights are $(w_1, w_2, w_3) = (0.8, 0.1, 0.1)$, reflecting the dominance of
the rank signal in production. Adaptive hard negative sampling uses top-$K=5$
candidates with concentration parameter $\eta=2.0$. Impression feedback weights
are $(\omega_{\text{imp}}, \omega_{\text{clk}}, \omega_{\text{cvr}}) = (0.5,
1.0, 2.0)$. All models are trained on 8 data-center GPUs with AdamW
($\beta_1=0.9$, $\beta_2=0.999$) and linear warmup over the first 5\% steps.

\vspace{2pt}
\noindent
\textbf{Online metrics.} For the A/B test, we report Module CTR, Search Volume, Search DAU, and
serving-side statistics (P50/P95 latency and incremental resource cost).

\subsection{Baselines}
\label{subsec:baselines}

We compare against production-strength retrieval systems and recent generative query recommendation methods.

\vspace{2pt}
\noindent
\textbf{Industrial non-generative baselines.}
\textbf{SwingQ2Q}~\cite{cite_swing}: graph-based query-to-query
recommendation trained via the Swing algorithm on user search co\-occurrence.
\textbf{Pop/Trend}~\cite{cite_pop_recommendation}: time-aware and region-aware trending query recommendation.
\textbf{Dense-Ret}~\cite{cite_dssm, cite_youtube_dnn}: a dual-encoder retrieval model over a fixed query pool with user and context features.

\vspace{2pt}
\noindent
\textbf{Generative baselines.} \textbf{ZS-LLM}: zero-shot prompting of Qwen3.5-122B-A10B with user
behavior, GPS, and temporal context, generating $K$ queries without
fine-tuning.
\textbf{GQR}~\cite{guo2025gqr}: LLM-based generative query recommendation
with CTR-alignment training.
\textbf{Local-SUG}~\cite{chen2026localsug}: a geography-aware LLM for query
suggestion in local-life services with city-aware candidate mining, the most
task-aligned academic baseline.
\textbf{CTR-GQS}~\cite{cite_gqs}: business-aware generative baseline for
conversational search.
\textbf{OneSug}~\cite{cite_onesug}: end-to-end generative framework for
e-commerce query suggestion.
\textbf{AIGQ}~\cite{cite_aigq}: the closest industrial baseline; listwise training, IL-GRPO policy optimization, hybrid nearline/online deployment.

\vspace{2pt}
For fair comparison, all fine-tuned generative baselines are re-implemented and retrained from scratch on the same 30-day corpus under their original training objectives; ZS-LLM is evaluated zero-shot without fine-tuning. All methods receive the same input features, including GPS coordinates, temporal context, user profile, and behavior history. The difference lies only in feature encoding: AIGQ, OneSug, and CTR-GQS represent location and time as text tokens, while SPADE uses continuous multi-scale spatiotemporal encodings. For cross-domain methods (e.g., OneSug), we apply minimal adaptations by replacing item vocabularies with query text while preserving the original loss formulation.

\subsection{Main Offline Comparison (RQ1)}
\label{subsec:main_results}

Table~\ref{tab:main} presents the main offline results. SPADE outperforms all baselines in relevance (HR@5 $+5.6\%$, MRR@5 $+5.2\%$), executability (Exec@5 $+1.9\%$), and geographic grounding (GeoC@5 $-40.3\%$). The one metric where SPADE does not lead is diversity: IDiv@5 (0.211) is $3.4\%$ higher than AIGQ (0.204), i.e., marginally less diverse. The gap is structural: AIGQ's listwise optimization conditions each candidate on previously generated outputs, giving it a natural deduplication advantage, but relies on a ranking model to judge list quality, a noisy signal that weakens relevance. SPADE's pointwise alignment (MARS-DPO) provides more accurate per-candidate rewards at the cost of cross-candidate visibility during training. SDBS recovers most of the diversity loss at decoding time---without it, IDiv@5 would degrade to 0.231 ($+13.2\%$ over AIGQ)---and as Figure~\ref{fig:diversity} shows, SDBS achieves a strictly better relevance--diversity Pareto frontier than Diverse Beam Search. The remaining gains in relevance and grounding stem primarily from continuous spatiotemporal encoding (STAR) and business-aligned tokenization (OA-BPE).

For reference, non-generative baselines like SwingQ2Q achieve competitive relevance but lack diversity due to reliance on historical query co-occurrence. Among generative baselines, OneSug slightly outperforms Local-SUG in relevance, while Local-SUG benefits from location modeling in GeoC@5 and Exec@5. GQR surpasses CTR-GQS in relevance, whereas CTR-GQS improves diversity through its alignment objective. AIGQ is the strongest overall baseline, yet SPADE surpasses it across the key metrics.

\begin{table}[t]
  \centering
  \caption{Main offline comparison ($K=5$). Best in \textbf{bold};
  second-best \underline{underlined}. HR is reported in percentage points.
  $\dagger$: significant over AIGQ
  ($p < 0.01$, paired $t$-test over per-user metrics). $\downarrow$: lower is better.
  Sim = SemSim, Div = IDiv, Geo = GeoC.}
  \label{tab:main}
  \small
  \setlength{\tabcolsep}{2.5pt}
  \begin{tabular}{l ccc c cc}
    \toprule
    & \multicolumn{3}{c}{\textit{Relevance}} & \textit{Div.}
    & \multicolumn{2}{c}{\textit{Grounding}} \\
    \cmidrule(lr){2-4}\cmidrule(lr){5-5}\cmidrule(lr){6-7}
    Method & HR & Sim & MRR & Div$\downarrow$ & Geo$\downarrow$ & Exec \\
    \midrule
    Pop/Trend   & 6.1  & .256 & .0423 & .318 & 2.73 & .784 \\
    Dense-Ret   & 8.3  & .298 & .0614 & .291 & 2.48 & .841 \\
    SwingQ2Q    & 9.7  & .324 & .0738 & .279 & 2.31 & .897 \\
    \midrule
    ZS-LLM      & 5.4  & .241 & .0387 & .237 & 3.52 & .793 \\
    CTR-GQS     & 10.1 & .342 & .0782 & .245 & 2.94 & .887 \\
    GQR         & 10.8 & .356 & .0836 & .251 & 2.83 & .871 \\
    OneSug      & 11.3 & .367 & .0891 & .234 & 3.05 & .903 \\
    Local-SUG   & 11.0 & .362 & .0867 & .241 & \underline{2.28} & .915 \\
    AIGQ        & \underline{12.5} & \underline{.386} & \underline{.1012} & \textbf{.204} & 2.78 & \underline{.934} \\
    \midrule
    \textbf{SPADE} & \textbf{13.2}$^\dagger$ & \textbf{.398}$^\dagger$ & \textbf{.1065}$^\dagger$
                   & \underline{.211}
                   & \textbf{1.66}$^\dagger$ & \textbf{.952}$^\dagger$ \\
    \midrule
    \textit{$\Delta$ AIGQ} & \small+5.6\% & \small+3.1\% & \small+5.2\% & \small+3.4\% & \small$-$40.3\% & \small+1.9\% \\
    \bottomrule
  \end{tabular}
\end{table}

\subsection{Component Ablation (RQ2)}
\label{subsec:ablation}

Table~\ref{tab:ablation} reports the ablation study, with five variants:
(1)~\textbf{w/o OA-BPE}: replace OA-BPE with standard BPE;
(2)~\textbf{w/o STAR}: replace the spatiotemporal encoder with a coarse
discrete location token;
(3)~\textbf{w/ Std-DPO}: replace MARS-DPO with standard DPO (uniform negative
sampling, equal sample weights);
(4)~\textbf{w/ Std-BS}: replace Sigmoid Diversity Beam Search with standard
beam search;
(5)~\textbf{w/ DBS}: replace Sigmoid Diversity Beam Search with Diverse Beam
Search \cite{cite_beam_search_degeneration}.

Every component contributes positively. OA-BPE yields the largest HR@5 gain
(12.5 $\rightarrow$ 13.2, +5.6\%), confirming that business-aligned tokenization improves the generation
prior for high-value long-tail queries; its effect is localized to relevance,
with IDiv@5 and GeoC@5 nearly unchanged. Removing STAR degrades all four
metrics: without fine-grained spatiotemporal encoding, the model cannot capture
interests tied to the user's immediate surroundings or time-sensitive demand
patterns, reducing both relevance ($-$4.5\% HR@5) and diversity (+5.7\%
IDiv@5) while causing the largest GeoC@5 increase (+53.6\%). MARS-DPO improves
HR@5 by 3.1\% and MRR@5 by 3.5\% over standard DPO, confirming that adaptive
hard negative sampling and impression feedback weighting strengthen preference
alignment. Sigmoid Diversity Beam Search reduces IDiv@5 by 8.7\% over standard
beam search with negligible impact on HR@5 ($<$0.2\%). DBS achieves
slightly better IDiv@5 (0.208) than SDBS, but its forced beam grouping
generates low-relevance candidates, causing a $-$34.1\% HR@5 collapse and
severe GeoC@5 degradation (2.82 km), a trade-off that is unacceptable in production.

\begin{table}[t]
  \centering
  \caption{Ablation study. All variants share the same backbone and training
  data. Best results are \textbf{bold}. $\downarrow$: lower is better.}
  \label{tab:ablation}
  \small
  \begin{tabular}{l cccc}
    \toprule
    Variant & HR@5 & MRR@5 & IDiv@5$\downarrow$ & GeoC@5$\downarrow$ \\
    \midrule
    w/o OA-BPE      & 12.5 & 0.1024 & 0.213 & 1.68 \\
    w/o STAR        & 12.6 & 0.1018 & 0.223 & 2.55 \\
    w/ Std-DPO      & 12.8 & 0.1029 & 0.219 & 1.71 \\
    w/ Std-BS       & 13.2 & 0.1062 & 0.231 & 1.67 \\
    w/ DBS          & 8.7  & 0.0651 & \textbf{0.208} & 2.82 \\
    \midrule
    \textbf{SPADE}  & \textbf{13.2} & \textbf{0.1065} & 0.211 & \textbf{1.66} \\
    \bottomrule
  \end{tabular}
\end{table}

\subsection{Slice Analysis for Local Search (RQ3)}
\label{subsec:slice}

Table~\ref{tab:slice} reports HR@5 and GeoC@5 on four traffic slices,
and SPADE's advantage over AIGQ grows in the most difficult conditions. On
cold-start users (fewer than 5 interactions), SPADE achieves +9.8\% HR@5,
suggesting that OA-BPE and STAR provide better inductive bias when user history
is sparse. On long-tail intents (query frequency $<$100 in the training log),
SPADE gains +13.5\% HR@5, consistent with OA-BPE's intended effect: more favorable
tokenizations for business-valuable long-tail queries shift the generation prior
toward rare but important intents. On geographically sparse regions (fewer than
10K monthly active users), SPADE reduces GeoC@5 by 48.3\% (lower is better),
confirming that continuous spatiotemporal encoding is most valuable when discrete
location tokens are too coarse to capture fine-grained regional demand. On
time-sensitive traffic (weekday meal hours 11--13 and 17--20), both methods
achieve higher HR@5 because weekday dining near the workplace is high-frequency
and predictable; the GeoC@5 gap narrows to $-$28.1\% as coarse location
signals already capture routine office-area demand.

\begin{table}[t]
  \centering
  \caption{Slice analysis. HR@5 and GeoC@5 (km, $\downarrow$) for SPADE and AIGQ
  on four traffic subsets. $\Delta$: relative change of SPADE over AIGQ.}
  \label{tab:slice}
  \small
  \setlength{\tabcolsep}{3.5pt}
  \begin{tabular}{l cc cc}
    \toprule
    & \multicolumn{2}{c}{HR@5} & \multicolumn{2}{c}{GeoC@5$\downarrow$} \\
    \cmidrule(lr){2-3}\cmidrule(lr){4-5}
    Slice & AIGQ & SPADE ($\Delta$) & AIGQ & SPADE ($\Delta$) \\
    \midrule
    Cold-start       & 9.2  & 10.1 (+9.8\%)   & 3.41 & 1.94 ($-$43.1\%) \\
    Long-tail        & 7.4  &  8.4 (+13.5\%)  & 3.58 & 2.07 ($-$42.2\%) \\
    Geo-sparse       & 10.8 & 11.6 (+7.4\%)   & 4.12 & 2.13 ($-$48.3\%) \\
    Time-sensitive   & 16.8 & 17.9 (+6.5\%)   & 1.92 & 1.38 ($-$28.1\%) \\
    \bottomrule
  \end{tabular}
\end{table}

\subsection{Analysis of OA-BPE and STAR (RQ2 cont.)}
\label{subsec:component_analysis}

\vspace{2pt}
\noindent
\textbf{OA-BPE tokenization quality.}
Table~\ref{tab:oabpe} compares OA-BPE with standard BPE across head queries
(frequency $\geq$1000) and long-tail queries (frequency $<$100). On head
queries, OA-BPE produces nearly identical tokenizations ($-$2.4\% median
length), confirming that business-value weighting does not degrade
high-frequency terms. On long-tail queries, OA-BPE reduces the median token
length by 18.3\% and increases their generation probability in the top-10
output list by 22.7\%. The fraction of long-tail queries whose tokenization is
shortened (by at least one token) reaches 64.1\%, indicating that the
OA-BPE merge criterion systematically favors commercially valuable rare
phrases.

\begin{table}[t]
  \centering
  \caption{OA-BPE vs.\ standard BPE tokenization statistics.}
  \label{tab:oabpe}
  \small
  \setlength{\tabcolsep}{4pt}
  \begin{tabular}{llccr}
    \toprule
    Subset & Metric & Std & OA & $\Delta$ \\
    \midrule
    \multirow{2}{*}{Head ($\geq$1K)}
      & Med.\ tok.\ len.       & 4.1   & 4.0   & $-$2.4\%  \\
      & Gen.\ prob.\ top-10    & 0.587 & 0.591 & +0.7\%    \\
    \midrule
    \multirow{4}{*}{Tail ($<$100)}
      & Med.\ tok.\ len.       & 7.2   & 5.9   & $-$18.3\% \\
      & Gen.\ prob.\ top-10    & 0.312 & 0.383 & +22.7\%   \\
      & Frac.\ shortened       & ---   & ---   & 64.1\%    \\
      & Avg.\ toks saved       & ---   & ---   & 1.8       \\
    \bottomrule
  \end{tabular}
\end{table}

\vspace{2pt}
\noindent
\textbf{STAR spatiotemporal encoding.}
Table~\ref{tab:star_ablation} isolates the contribution of each STAR component.
We evaluate five variants:
(1)~\textbf{w/o spatial}: removes the multi-scale spatial sinusoidal encoding
and relies solely on temporal signals and user history;
(2)~\textbf{w/o temporal}: removes the periodic temporal encoding (hour-of-day
and day-of-week) and relies solely on spatial signals;
(3)~\textbf{single-scale}: replaces the multi-scale spatial encoding
($S=4$ resolutions from 1\,km to city-level) with a single 1\,km resolution,
removing the coarser scales that capture regional demand patterns;
(4)~\textbf{no gate}: replaces the zero-initialized soft gate
$\alpha \cdot \mathbf{e}_{st}$ with direct addition of spatiotemporal
embeddings to the hidden state, removing the gradual warm-up mechanism;
(5)~\textbf{SPADE (full)}: the complete STAR module with multi-scale spatial
encoding, periodic temporal encoding, and soft gating.

\begin{table}[t]
  \centering
  \caption{STAR component ablation. All variants use the same OA-BPE
  vocabulary, MARS-DPO training, and SDBS decoding. $\downarrow$: lower is
  better.}
  \label{tab:star_ablation}
  \small
  \begin{tabular}{l cccc}
    \toprule
    Variant & HR@5 & MRR@5 & IDiv@5$\downarrow$ & GeoC@5$\downarrow$ \\
    \midrule
    w/o spatial       & 12.7 & 0.1031 & 0.213 & 2.55 \\
    w/o temporal      & 13.0 & 0.1048 & 0.212 & 1.79 \\
    single-scale      & 12.9 & 0.1041 & 0.214 & 1.91 \\
    no gate           & 12.8 & 0.1033 & 0.216 & 1.82 \\
    \midrule
    \textbf{SPADE (full)} & \textbf{13.2} & \textbf{0.1065} & \textbf{0.211} & \textbf{1.66} \\
    \bottomrule
  \end{tabular}
\end{table}

Removing spatial encoding causes the largest GeoC@5 degradation
(2.55\,km, +53.6\%) and the largest HR@5 drop ($-$3.8\%), confirming that
geographic grounding depends primarily on multi-scale spatial signals and that
spatial context also contributes to intent relevance. Removing temporal
encoding has a smaller effect on GeoC@5 (1.79\,km, +7.8\%) but still hurts
HR@5 ($-$1.5\%), indicating that time-of-day patterns contribute to relevance
by capturing periodic demand shifts (e.g., lunch vs.\ evening intents). The
single-scale variant degrades both GeoC@5 (1.91\,km) and HR@5 (12.9), showing
that coarser scales encode complementary regional demand that a single fine
resolution cannot capture alone. Removing the soft gate yields HR@5 of 12.8
($-$3.0\%) despite retaining spatial and temporal signals: directly injecting
spatiotemporal embeddings into the hidden state disrupts pre-trained language
model representations during early training, whereas the zero-initialized gate
lets the model incorporate these signals gradually as training stabilizes.

GeoC@5 also decreases monotonically as the finest spatial resolution is
refined: 2.68\,km with city-level tokens (close to the AIGQ baseline at
2.78\,km), 2.41\,km at 10\,km, 1.98\,km at 5\,km, and 1.66\,km at 1\,km.

\subsection{Diversity Analysis}
\label{subsec:diversity_analysis}

\begin{figure}[htbp]
  \centering
  \includegraphics[width=\linewidth]{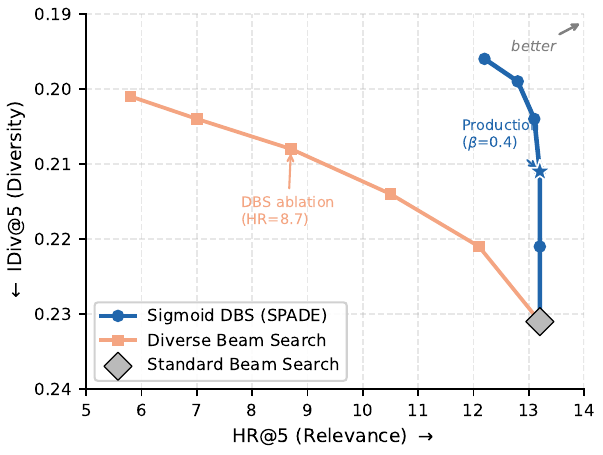}
  \Description{Pareto frontier plot of HR@5 versus IntraDiv@5 as beta
  varies, comparing Sigmoid Diversity Beam Search, Diverse Beam Search,
  and standard beam search.}
  \caption{Relevance--diversity trade-off as diversity strength varies.
  The $y$-axis is inverted so that the upper-right corner is optimal
  (high relevance, high diversity). SDBS (SPADE) achieves a strictly
  better Pareto frontier than DBS: at comparable IDiv@5, SDBS retains
  near-full HR@5 while DBS collapses to 8.7 HR@5 ($-$34\%).}
  \label{fig:diversity}
\end{figure}

Figure~\ref{fig:diversity} shows the relevance--diversity trade-off as the
bonus strength $\beta$ varies from 0 to 1. At $\beta=0$, all methods reduce
to standard beam search. As $\beta$ increases, Sigmoid Diversity Beam Search
achieves a strictly better Pareto frontier than Diverse Beam Search
\cite{cite_beam_search_degeneration}: at the same HR@5, SPADE's decoder yields
lower IDiv@5 (more diverse), and vice versa. The production operating point
($\beta=0.4$) corresponds to the results in Table~\ref{tab:main}.

\subsection{Online A/B Test (RQ4)}
\label{subsec:online}

We deploy SPADE on the production platform and run a one-week A/B test against
the production baseline with 20\% of traffic per arm.
SPADE achieves +1.88\% relative lift in Module CTR ($p < 0.05$), +0.177\% in Search Volume
($p < 0.05$), and +0.105\% in Search DAU ($p < 0.05$), indicating that offline
gains transfer to production.

\textbf{Serving performance.} On a single inference-class GPU, SPADE occupies
about 20\,GB of memory and achieves a throughput of 289 QPS with P50 latency of
36.1\,ms and P95 latency of 65.4\,ms. This satisfies the platform's real-time serving constraint
(P95 $<$ 100\,ms) with comfortable headroom, and the per-request cost is
comparable to the previous generation model, requiring no additional GPU
allocation for deployment.

\section{Conclusion}
In this work, we presented SPADE, an end-to-end generative framework for proactive query recommendation in local search. Its four components---OA-BPE, STAR, MARS-DPO, and Sigmoid Diversity Beam Search---form a single co-designed pipeline spanning the vocabulary, context encoding, preference alignment, and decoding stages. Offline experiments show +5.6\% HR@5 and $-$40.3\% GeoC@5 over the strongest baseline; a one-week online A/B test confirms the offline results, with +1.88\% Module CTR, +0.177\% Search Volume, and +0.105\% Search DAU. Limitations point to future directions: the quality reward relies on Qwen3.5-122B-A10B batch inference, which a distilled reward model could replace; OA-BPE weights require periodic recomputation as the query distribution evolves; and STAR encodes a single GPS coordinate, leaving trajectory-level spatiotemporal context as a natural extension.

 \end{spacing}

\bibliographystyle{ACM-Reference-Format}
\bibliography{references}

\end{document}